\documentstyle[epsfig,aps]{revtex}

\begin{document}

\draft

\def\lsim{\lower.5ex\hbox{$\; \buildrel < \over \sim \;$}}
\def\gsim{\lower.5ex\hbox{$\; \buildrel > \over \sim \;$}}


\title{Space-time curvature coupling of spinors in early universe: Neutrino asymmetry and 
a possible source of baryogenesis} 
\author{Ujjal Debnath$^1$\thanks{ujjal@iucaa.ernet.in}, 
Banibrata Mukhopadhyay$^2$\thanks{bmukhopa@cfa.harvard.edu}, 
Naresh Dadhich$^3$\thanks{nkd@iucaa.ernet.in} 
\\ \vskip0.5cm
{\sl 1. Department of Mathematics, Jadavpur University, Kolkata-700032, India}\\
{\sl 2. Theory Division, Harvard-Smithsonian Center for Astrophysics, 60 Garden Street,
MS-51, Cambridge, MA 02138, USA}\\
{\sl 3. Inter-University Centre for Astronomy and Astrophysics, Post Bag 4,\\ 
Ganeshkhind, Pune-411007, India}
}

\maketitle
\baselineskip = 18 true pt
\vskip0.3cm
\setcounter{page}{1}

\def\ch{\lower-0.55ex\hbox{--}\kern-0.55em{\lower0.15ex\hbox{$h$}}}
\def\lh{\lower-0.55ex\hbox{--}\kern-0.55em{\lower0.15ex\hbox{$\lambda$}}}       
\def\n{\nonumber}

\begin{abstract}

It is well known that when a fermion propagates in curved
space-time, its spin couples to the
curvature of background space-time. We propose that this interaction for neutrinos
propagating in early curved universe could give rise to
a new set of dispersion relations and then neutrino asymmetry at equilibrium.
We demonstrate this with the Bianchi models which describe the homogeneous but anisotropic
and axially symmetric universe. If the lepton number violating processes freeze out at $10^{-37}$
second when temperature $T\sim 10^{15}$GeV, neutrino asymmetry of the order of 
$10^{-10}$ can be generated. A net baryon asymmetry of the same magnitude can thus be
generated from this lepton asymmetry either by a GUT $B-L$ symmetry or
by the electro-week sphaleron processes which have $B+L$ symmetry.

\end{abstract}

\vskip1.0cm
\pacs{KEY WORDS :\hskip0.3cm neutrino asymmetry, early universe, anisotropic space-time, CPT 
violation 
\\
\vskip0.1cm
PACS NO. :\hskip0.3cm 04.62.+v, 11.30.Er, 11.30.Fs  }
\vskip2cm



The asymmetry of neutrino in universe is a known fact. This asymmetry is thought to
arise due to lepton number asymmetry, e.g., via the Affleck-Dine  mechanism \cite{admcdonald}.
There are several important consequences of a large neutrino asymmetry in early universe
which may have effects on production of light elements during BBN, contribution of relic 
neutrinos to the present energy density of universe, change of neutrino decoupling temperature,
cosmic microwave background etc. \cite{cosmoeffects}.
Also the massive neutrinos with large asymmetry can explain the existence
of cosmic radiation with energy greater than GZK cutoff \cite{gzk}.
Keeping all these importance in mind, our present goal is to prescribe a new mechanism
which can give an insight to the origin of neutrino asymmetry fixed up in the early era. 

When any fermion propagates in curved space-time its
spin couples to the background curvature connection and gives rise to an
interaction. The various aspects of this feature have been studied in past 
(e.g., \cite{pap51,p71,sw,dixon,chandra76,deb,an94,l95,m00}). It has also been shown that this
interaction may be responsible for an additional neutrino asymmetry even in the present era
of universe \cite{ms03,sm03,m05}. However, it is very difficult to visualize 
as the strength of the above mentioned interaction is negligible in our earth 
which is practically flat. In a similar 
fashion, the space-time around a black hole
can generate neutrino asymmetry locally \cite{ms03,m05}, but as we do not have overall information
about number of black holes and their corresponding spin orientation, it is very difficult to
predict whether it produces a significant asymmetry over the relic value
or not. A similar mechanism to produce asymmetry was noted earlier \cite{ahl1}. 
Later, the Lorentz and the CPT violating scenarios were
addressed in the context of Riemann-Cartan space-times \cite{kosgrav} and 
in the neutrino sector
\cite{km}, although in curved space-time a precise definition of the 
CPT symmetry is challenging to establish. 

The important fact to note is that the interaction term, arising due to spin-curvature 
coupling, does not preserve CPT and is similar in its mathematical form 
to the effective CPT violating
terms known to exist in other contexts (see, e.g. \cite{ck,d99,bklr}).
This interaction has opposite sign for a neutrino and an anti-neutrino, and therefore 
splits up their energy levels which may violate lepton symmetry in a certain situation. 
Therefore, if neutrinos are considered to be
propagating in the non-flat early universe era, then due to presence of lepton
number violating GUT processes a net asymmetry may arise between neutrinos
and anti-neutrinos at the thermodynamic equilibrium. With time, as temperature goes down,
this neutrino asymmetry also goes down and gets frozen-in when the lepton number
violating GUT processes decouple at the era when temperature, $T\lsim 10^{15}$GeV. 
Note that 
according to the Linde's chaotic inflation model \cite{linde}, inflation would start at the Planck time
and end at the era when temperature, $T\sim 10^{14}-10^{15}$GeV. On the other hand, if inflation would
start at the post Planck time \cite{guth}, then again it would end at similar temperature. 
Therefore in either way, there is a minimum chance to
wipe out this neutrino asymmetry at the end of inflation. Moreover, this neutrino asymmetry may
be favored during inflation, i.e. in presence of primordial quantum fluctuations in the 
space-time. This is 
basically the tensor perturbation to early universe which also brings the off-diagonal
terms in the metric responsible for the CPT violation as mentioned above. Therefore,
one could argue for the gravity wave induced neutrino asymmetry in early universe.

If there is
a chance to wipe out this asymmetry during inflation, the space-time curvature effect would still
split up the energy of a neutrino from that of an anti-neutrino, which might give rise to an additional asymmetry
solely due to the curvature effect of early universe. In presence of gravity,
origin of this CPT violating interaction is an interesting result on its own right. 
The magnitude of neutrino asymmetry
depends on the order of anisotropy as well as the time when the lepton number violating
processes freeze out in the early era. Here we 
show that the generated neutrino asymmetry by our mechanism 
agrees with observation perfectly.
The basic criteria to generate neutrino asymmetry in early universe through our
mechanism are: (i) The space-time of early universe should have deviated from spherical symmetry.
(ii) The interaction Dirac Lagrangian must be CPT violating, at least in a 
local frame, which may be an
axial four-vector (or pseudo four-vector) multiplied by a curvature coupling four-vector potential.
(iii) The temperature scale of the system should be large with respect
to the energy scale of the space-time curvature.

Therefore we show that one of the possible origin of neutrino asymmetry is the anisotropic phase
of early universe. 
As the background metric deviates from spherical symmetry, 
neutrino asymmetry comes out clearly. In this connection, the Dirac equation and the 
corresponding Lagrangian come into the picture for obvious reason. 
One of the key requirement to generate neutrino asymmetry by this mechanism 
is that the background metric should have at least one off-diagonal spatial component, 
if the set of coordinate variables is $\{x,y,z,t\}$.
If early universe is thought to be anisotropic, we achieve the required
form of metric. On the other hand, as long as the anisotrope is low,
which is good enough for the present purpose,
 the new cosmological data of WMAP are quite consistent with an anisotropic universe.
Therefore, we consider a simplified version of Bianchi II, VIII and IX models \cite{bsc}.
The generalized form of the metric is

\begin{equation}
d s^2 = -dt^2+S(t)^2\, dx^2+R(t)^2\,[dy^2+f(y)^2\,dz^2]-S(t)^2\,h(y)\,[2dx-h(y)\,dz]\,dz
\label{mat}
\end{equation}
where for the Bianchi II, VIII and IX models, respectively $f(y)$ and $h(y)$ are given as
\begin{equation}
f(y)=\{y,\sinh y,\sin y\},\hskip1cm h(y)=\{-y^2/2,-\cosh y,\cos y\}.
\end{equation}
                                                                                                                                 
The corresponding orthogonal set of non-vanishing components of tetrad (vierbein) can be written as
\begin{eqnarray}
\nonumber
&&e^0_t = 1,\,\,e^1_x=f(y)R(t)S(t)/\sqrt{f(y)^2R(t)^2+S(t)^2h(y)^2},\,\,e^2_y=R(t),\\
&&e^3_z=\sqrt{f(y)^2R(t)^2+S(t)^2h(y)^2},\,\,e^3_x=-S(t)^2h(y)/\sqrt{f(y)^2R(t)^2+S(t)^2h(y)^2}.
\label{tetnonv}
\end{eqnarray}

Thus the generalized Dirac Lagrangian density in absence of torsion
can be given as
\begin{equation}
{\cal L}=\sqrt{-g}\left(i \, \bar{\psi} \, \gamma^aD_a\psi- m \, \bar{\psi}\psi\right),\label{lag}
\end{equation}
where the covariant derivative and spin connection are defined as
\begin{eqnarray}
D_a=\left(\partial_a-\frac{i}{4}\omega_{bca}\sigma^{bc}\right) \label{cd},
\end{eqnarray}

\begin{eqnarray}
\omega_{bca}=e_{b\lambda}\left(\partial_a e^\lambda_c+\Gamma^\lambda_{\gamma\mu} e^\gamma_c e^\mu_a\right). \label{om}
\end{eqnarray}
We work in units of $c = \hbar = k_B  = 1$. The Lagrangian is invariant under local Lorentz
transformation of vierbein and spinor field as
$e^a_\mu(x) \rightarrow \Lambda^a_b(x) e^b_\mu(x)$ and $\psi(x) \rightarrow \exp(i \epsilon_{a b}(x) \sigma^{a b}) \psi(x)$, where $\sigma^{ab}=\frac{i}{2}\left[\gamma^a,\gamma^b\right]$, is the 
generator of tangent space Lorentz transformation. The Latin and the Greek alphabets indicate
the flat and the curved space coordinates respectively. Also
\begin{eqnarray}
e^\mu_a e^{\nu a}=g^{\mu\nu},\hskip0.5cm e^{a \mu} e^b_\mu=\eta^{ab},\hskip0.5cm \{\gamma^a,\gamma^b\}=2\eta^{ab},
\end{eqnarray}
where $\eta^{ab}$ represents the inertial frame of the Minkowski metric and $g^{\mu\nu}$ the curved space-time metric. 

If we expand eqn. (\ref{lag}), spin connection terms are reduced to the 
combination of an anti-hermitian,
$\bar{\psi} A_a\gamma^a \psi$, and a hermitian, $\bar{\psi} B^d \gamma^5 \gamma_d \psi$, terms \cite{mmp02}.
The anti-hermitian interaction term disappears when its conjugate part is added to the Lagrangian.
The only interaction survives in $\cal L$ is the hermitian part and then eqn. 
(\ref{lag}) reduces to
\begin{eqnarray}
{\cal L}={\cal L}_f+{\cal L}_I=\sqrt{-g}\bar{\psi}\left[(i\gamma^a\partial_a-m)+\gamma^a\gamma^5 B_a\right]\psi,
\label{lagf}
\end{eqnarray}
where
\begin{eqnarray}
B^d=\epsilon^{abcd} e_{b\lambda}\left(\partial_a e^\lambda_c+\Gamma^\lambda_{\alpha\mu}
e^\alpha_c e^\mu_a\right)\label{bd}.
\label{bd}
\end{eqnarray}
The explicit form of gravitational scalar potential, $B_0$ (which is the most important 
quantity in our formalism that we show later), can be written as
\begin{eqnarray}
 B^0 &=& e_{1\lambda}  \left( \partial_3 e_2^\lambda - \partial_2 e_3^\lambda
 \right) + e_{2 \lambda} \left( \partial_1 e_3^\lambda - \partial_3 e_1^\lambda
 \right) + e_{3 \lambda} \left( \partial_2 e_1^\lambda - \partial_1 e_2^\lambda
 \right).
\label{bo}
\end{eqnarray}
Similarly, gravitational vector potentials, $B^1,B^2,B^3$, can be evaluated.
From eqn. (\ref{bo}), it is clear that $B_0$ is zero if all the
off-diagonal spatial components of the metric are zero (i.e. $g_{ij}=0$,
where, $i\neq j\rightarrow 1,2,3$). 

In eqn. (\ref{lagf}), the free part of the Lagrangian 
is same as the Dirac Lagrangian in flat space, except the  
multiplicative factor 
$\sqrt{-g}$ which is unity in flat space. The interaction part is an axial-vector
multiplied by a gravitational four-vector potential. We know that the Lagrangian for 
any fermionic field is 
invariant under the local Lorentz transformation \cite{bd}. However, if the background gravitational field, $B_a$, 
is chosen to be constant in the local frame, then 
${\cal L}_I$ violates CPT as well as the particle Lorentz symmetry in 
the local frame. For example, if $B_a$ is constant and space-like, then the corresponding fermion will have different
interaction if its direction of motion or spin orientation changes, and thus results
the breakdown of
Lorentz symmetry in the local frame. This is the key conception of our present formalism.
Similar interaction terms are considered in CPT violating theories and string theory (e.g. \cite{ck}, \cite{kp}),
but in the present case these originate automatically. The interaction 
${\cal L}_I$ is observer Lorentz invariant but violates the particle Lorentz symmetry
(see, e.g. \cite{ck,bd}).
If ${\cal L}_I$ changes sign under the CPT transformation, then we understand that
it does not preserve CPT. Actually, under the CPT transformation,
associated axial-vector or pseudo-vector ($\bar{\psi}\gamma^a\gamma^5\psi$) changes sign.
Now, as $B_a$ is a constant coupling in the local frame, ${\cal L}_I$ violates CPT. 
If $B_a$ is treated as a background field in a local
frame, then the interaction violates CPT explicitly. When there is no 
back-reaction of the microphysics involving the fermions on the metric 
and $B_a$ is considered as a fixed external field, then CPT is violated 
spontaneously.
However, in the present case, with its functional form we can determine 
the explicit CPT status of $B_a$ itself along the space-time. 
If $B_a(x,y,z,t)$ is not an odd function under CPT 
[$B_a(-x,-y,-z,-t)\neq - B_a(x,y,z,t)$], then ${\cal L}_I$ comes out to be
a CPT violating (CPT odd) interaction along the space-time. 
The nature of background metric determines
whether $B_a(x,y,z,t)$ is odd under CPT or not. Overall we can say, ${\cal L}_I$ is CPT as well as particle
Lorentz violating interaction. 

It can be noted in this respect that, assuming implicitly all 
fields are standard model fields, CPT violation necessarily implies the Lorentz
violation in local field theory \cite{greenberg}. However, this is not valid for other Wigner classes 
\cite{wigner} for which the Lee-Wick theorem \cite{lee} assures non-locality. 
Recently, a new form of CPT violation has been shown \cite{ag1,ag2} 
that arises without violating the Lorentz symmetry.

In our case, the four-vector $B_a$ is treated as a Lorentz-violating
and CPT-violating spurion. However, if $B_a$ does not break the symmetry of particle Lorentz transformations 
in the local frame, then the CPT preserves.  We plan to show that the
fermion propagating in early universe governs the CPT violating interaction.
It was shown in an earlier work \cite{mmp02} that 
the space-time metric could be such that the $B_a(x,y,z,t)$ 
is CPT odd along the space-time and therefore the overall
interaction could be CPT invariant.

It is important to note that the interaction in the present formalism 
is different from those studied earlier which were also Lorentz violating but mainly
CPT even \cite{colglash}. Those studies were based on interactions
in present universe which is flat one and thus excluded the interactions
of fermions with background curvature.
The purpose of those studies was to have high energy and high precision tests of special relativity.
One could then obtain the bound on terms in the Lagrangian violating Lorentz invariance
through various experiments, like cosmic ray observations, neutrino
oscillations etc. We, in the present paper, concentrate on different aspects
and establish that the background
curvature plays an interesting role in disguise of vector $B_a$
to cause CPT violation and hence neutrino$-$anti-neutrino asymmetry in presence 
of lepton number violating process. 
As applied to the phenomenology, our motivation is to seek the possible origin of
neutrino$-$anti-neutrino asymmetry in early universe by putting bounds on
parameters. It would be interesting
to extend this analysis to study the phenomenological applications,
e.g., neutrino oscillation as studied earlier \cite{colglash}.

Thus the corresponding dispersion relations for left and right chiral fields (here the neutrino and the anti-neutrino)
are given by
\begin{eqnarray}
(p_a \pm B_a)^2=m^2,
\label{dis}
\end{eqnarray}
where the upper sign corresponds to particle and the lower sign to anti-particle.
Clearly the dispersion relation is modified due to presence of
the CPT violating term. 
Now, in the case of neutrino, we can identify left handed species as particle and right handed 
species as corresponding
anti-particle. Then, after some simple algebra, the energies for particle
($E_\nu$) and anti-particle ($E_{\bar{\nu}}$) are given by
\begin{eqnarray}
\nonumber
E_{\nu} &=& \sqrt{({\bf p} - {\bf B})^2+m^2}+B_0,\\
E_{\bar{\nu}} &=& \sqrt{({\bf p} + {\bf B})^2+m^2}-B_0,
\label{edis}
\end{eqnarray}
which indicate that neutrino and anti-neutrino propagating in presence of the space-time curvature
have different energies. 
Thus, there is an energy gap between left handed and right handed species, which is
proportional to the interaction term $B_a p^a$. When $B_a \longrightarrow 0$, physically
the case of Robertson-Walker universe which is spherically symmetric,
this helicity energy gap disappears.
Therefore, the difference of their number density in early
universe, namely neutrino asymmetry, can be evaluated by the integral
\begin{eqnarray}
\Delta n=\frac{g}{(2\pi)^3} \int d^3 {\bf p}
\left[\frac{1}{1+\exp(E_{\nu}/T)}-\frac{1}{1+\exp(E_{{\bar{\nu}}}/T)}\right].
\label{fn}
\end{eqnarray}

If $B_0=0$, the integrand is an odd function of $\bf p$ and hence 
$\Delta n = 0$. To have a non-zero neutrino
asymmetry, $B_0$ must be non-zero whether $B_i$s ($i=1,2,3$) are present or not. 
This is the reason that the space-time metric
should have a non-zero off-diagonal spatial components for neutrino asymmetry to occur.

According to the Bianchi model (\ref{mat}), only $B_0$ and $B_2$ are non-zero given as
\begin{eqnarray}
B^0=\frac{S[-f^2R^2(hf^\prime R+Sh^\prime)+h^2S^2(hf^\prime R+Sh^\prime)+2fhRS(Rf^\prime-hh^\prime S)]}
{f^4R^4+f^2h^2R^2S^2}
\label{b0}
\end{eqnarray}
\begin{eqnarray}
B^2=\frac{h[-f^2R^2+2fRS+h^2S^2][RS^\prime-R^\prime S]}
{f^3R^4+fh^2R^2S^2}.
\label{b2}
\end{eqnarray}

Now for Bianchi II:                                                                                          
\begin{eqnarray}
\nonumber
B^0&=&\frac{4R^3S+3y^2R\,S^3-2y\,S^4}{8R^4+2y^2R^2S^2}\\
B^2&=&\frac{(4y\,R^2-8R\,S-y^3S^2)(R\,S^\prime-R^\prime S)}{8R^4+2y^2R^2S^2},
\label{bian2}
\end{eqnarray}
for Bianchi VIII:
\begin{eqnarray}
\nonumber
B^0&=&\frac{S[2\cosh^2y(\cosh 2y-3)R\,S^2-4\cosh^2y\,\sinh y\,S^3+4R^3\,\cosh^2y\,\sinh^2y-R^2S(5\,\sinh y+
\sinh 3y)}
{4(\cosh^2y\,\sinh^2y\,R^2S^2+R^4\sinh^4y)}\\
B^2&=&\frac{\cosh y(S^2\cosh^2y+2R\,S\,\sinh y-R^2\,\sinh^2y)(RS^\prime-R^\prime S)}
{\cosh^2y\,\sinh y\,R^2S^2+R^4\sinh^3y},
\label{bian8}
\end{eqnarray}
and for Bianchi IX:
\begin{eqnarray}
\nonumber
B^0&=&\frac{S[2\cos^2y(3-\cos 2y)RS^2-4\cos^2y\,\sin y\,S^3-4R^3\cos^2y\,\sin^2y+R^2S(5\sin y+\sin 3y)]}
{4(\cos^2y\,\sin^2y\,R^2S^2+R^4\sin^4y)}\\
B^2&=&\frac{\cos y(S^2\cos^2y+2RS\,\sin y-R^2\sin^2y)(RS^\prime-R^\prime S)}{\cos^2y\,\sin y\,R^2S^2+R^4\sin^3y}.
\label{bian9}
\end{eqnarray}
It is very clear from above that $B^0$ (and also $B^2$)
does not flip sign under space-inversion,
i.e. for $y\rightarrow -y$. Thus, it is {\it not} an odd function over the space-time for
any of the Bianchi models and the form of $B_0$ is such that
$B_0(-x,-y,-z,-t) \neq \pm B_0(x,y,z,t)$. Therefore, $B_a$ leads to CPT violation
at any point $(x,y,z,t)$.
As mentioned earlier, along the space-time 
the nature of $B_a$ under CPT totally depends on the 
background metric, the space-time, where the neutrino propagates.
See \cite{mmp02} where a space-time is chosen such that $B_0(-x,-y,-z,-t)=-B_0(x,y,z,t)$
and hence ${\cal L}_I$ is CPT invariant.
However, the present case, where the space-time is chosen of early universe,
brings an actual CPT violating situation into the picture.

The axial vector part of ${\cal L}_I$ for particle, $\psi$, and anti-particle, $\psi^c$, may be expressed as
\begin{equation}
\bar{\psi}\gamma^a \gamma^5 \psi= \bar{\psi}_R \gamma^a \psi_R -\bar{\psi}_L \gamma^a \psi_L,
\hskip0.7cm                                                                                                             
\bar{\psi}^c \gamma^a \gamma^5 \psi^c = (\bar{\psi}^c)_R \gamma^a (\psi^c)_R
-(\bar{\psi}^c)_L \gamma^a (\psi^c)_L.
\label{antipart}
\end{equation}
According to the standard model, a neutrino is left-handed and an 
anti-neutrino is right-handed. Therefore, in early universe, ${\cal L}_I$ for a neutrino and an 
anti-neutrino respectively reduce as
\begin{equation}
{\cal L}_I=-{\overline \psi}_L\gamma^a\psi_LB_a,
\hskip0.7cm
{\cal L}_I=({\overline \psi}^c)_R\gamma^a(\psi^c)_RB_a.
\label{anineu}
\end{equation}
In addition, for the Majorana neutrinos, above ${\cal L}_I$ turns out explicitly as
\begin{equation}
{\cal L}_I=\psi_L^\dagger\gamma^a\psi_L B_a,
\hskip0.7cm
{\cal L}_I=-{\psi_L^c}^\dagger\gamma^a\psi_L^c B_a
\label{neulc}
\end{equation}
for left-handed particle, $\psi_L$, and corresponding charge conjugated particle, $\psi_L^c$.
Thus eqns. (\ref{dis}) and (\ref{edis}) are true for the Bianchi model and the neutrino 
asymmetry comes out off eqn. (\ref{fn}).

Let us now consider specifically the Bianchi II model with the choice of
$S(t)={\rm arbitrary\,\,constant}=C_1$ \cite{bsc}.
Let us also consider, for simplicity, that the space-time curvature is such that 
${(B_0)}^2 << B_0$, i.e. only the first order curvature effect is important. 
Thus, in the ultra-relativistic regime, we obtain from eqn. (\ref{fn})
\begin{eqnarray}
\Delta n=\frac{g}{(2\pi)^2} \, T^3 \, \int_0^\infty \, \int_0^\pi \, 
\left[\frac{1}{1+ e^u \, e^{ B_0/T}}-\frac{1}{1+ e^u \, e^{-  B_0/T}}\right]
\, u^2 \, d\theta \, du
\label{fn1}
\end{eqnarray} 
where $u = |{\vec p}|/T$. Therefore 
\begin{equation}
\Delta n \sim g \, T^3 \, \left(\frac{B_0}{T} \right).
\end{equation}
As long as the lepton number violating processes are in thermodynamical equilibrium, 
$\Delta n$ decreases as temperature goes down upto the
decoupling limit ($T_d$) for the lepton number violating processes. 
Then the net lepton number (here neutrino asymmetry) to entropy density 
(which is given as $s\sim T^3$) remains constant after decoupling and is given as
\begin{eqnarray}
\Delta L (T<T_d) =\frac{\Delta n}{s}&\sim & \frac{B_0}{T_d}.
\label{lepasy1}
\end{eqnarray}
If the lepton number violating GUT processes decouple at 
$T_d\sim 10^{28}$K $\sim 10^{15}$ GeV, when the age of universe, 
$t\sim 10^{-37}$ second, then the scale factor at that time could be given by
$R(t)\propto 10^{-19}$ 
($R(t)= (C_2\,t-C_3)^{1/2}$, when $C_2,C_3$ are arbitrary constants).
Thus, we can obtain $B_0\sim 10^5$ GeV.
Therefore, from eqn. (\ref{lepasy1}), neutrino as well as 
lepton asymmetry comes out to be $10^{-10}$,
which matches perfectly with observation. 
In general, a formula for lepton asymmetry in early universe can be given by
\begin{eqnarray}
\Delta L (T<T_d)\sim 10^{-10}\left(\frac{B_0}{10^5 GeV}\right)\left(\frac{10^{15} GeV}{T_d}\right).
\label{lepasy3}
\end{eqnarray}

Therefore, we propose a new mechanism to generate neutrino as well as lepton asymmetry in
early universe. We have explicitly demonstrated this 
when neutrinos are considered to be propagating in a space-time of early universe. 
The only requirement to generate neutrino asymmetry in this mechanism is that the 
early universe metric 
should have at least a non-zero space-space cross term (i.e. the off-diagonal spatial component; 
$g_{ij}, i\ne j\rightarrow 1,2,3$) when
the set of space-time coordinate is $\{x,y,z,t\}$. 
It is seen that, in presence of any $g_{ij}$, the scalar potential part ($B_0$) of 
space-time coupling is non-zero which 
is actually responsible for neutrino asymmetry in universe.
If all $g_{ij}$s are zero, $B_0$ and hence $\Delta n$ vanish. 

An important point to note is that after a long time, the homogeneous and 
anisotropic Bianchi model reduces to the space-time of present universe which
is isotropic. This is easily understood from the corresponding form of shear scalar. 
For the Bianchi II model (which is mainly used for the calculations
of various parameters in this problem), the shear scalar is obtained as $\sigma^2 \sim 1/t^2$,
which reduces to zero as $t\rightarrow \infty$. Therefore, although universe starts with an
anisotropic phase, with the choice of anisotrope consistent with
WMAP, it restores the complete isotropy at later period and reduces to that of present universe.

Our mechanism essentially works in 
presence of a pseudo-vector term ($\bar{\psi}\gamma^a\gamma^5\psi$) multiplied by a background
curvature coupling ($B_a$).
This is the CPT and the particle
Lorentz violating term, which picks up an opposite sign in between neutrino
and anti-neutrino. 
Thus we propose, to generate neutrino asymmetry in early universe, all the following
criteria have to be satisfied simultaneously: (i) The space-time must {\it not} be spherically 
symmetric. (ii) The interaction Dirac Lagrangian must have a CPT violating term,
at least locally, which may be an
axial-vector (or pseudo-vector) multiplied by a curvature coupling four-vector potential.
(iii) The temperature scale of the system should be large with respect
to the energy scale of the space-time curvature.

The early universe is a favorable era when
all the above conditions would satisfy. It would be interesting to explore 
further theoretical and phenomenological consequences of the role of
background gravitational curvature for neutrinos, which might offer new
insights in the interplay of gravity and standard model interactions and
specially of neutrino physics.

An interesting consequence of this fact may be the following. As GUT has $B-L$ symmetry, due to asymmetry 
of $L$, 
a baryon ($B$) asymmetry may be generated
of the same magnitude and sign as of lepton (neutrino) asymmetry. On the other hand, 
$B+L$ conservation of the sphaleron may give rise to a baryon asymmetry of
same magnitude and sign as of lepton asymmetry generated in the GUT. 
Thus we may pinpoint about the baryogenesis in universe.
A class of explicit CPT violating terms in the Lagrangian, which can generate baryon asymmetry,
have been studied in \cite{bert}. However, in our case, the basic origin of this 
CPT violating interaction and its connection to baryon asymmetry are different and 
inherent. It can be noted that the inclusion of torsion does not alter our
basic result. The presence of torsion only modifies the form of
$B^d$ in eqn. (\ref{bd}) without affecting the underlying physics.

\vskip1cm
\noindent{\large\bf Acknowledgment}\\
Authors thank A. Banerjee, S. Chakraborty, V. A. Kosteleck\'y, S. Mohanty, P. B. Pal, 
and A. R. Prasanna for discussions and
suggestions and N. Afshordi, R. Crocker, and K. Kohri for
discussions, at various stages of the work.
UD thanks CSIR (Government of India) for providing a senior research fellowship.
BM acknowledges the partial support to this research by
NSF grant AST 0307433.

\end{document}